\documentclass[aps,prl,twocolumn,superscriptaddress,showpacs]{revtex4-2}
\usepackage{natbib}
\usepackage{hyperref}
\usepackage[dvips]{graphicx}
\usepackage{amssymb,amsmath,amsfonts}
\usepackage[T1]{fontenc}
\usepackage{float}
\usepackage{color,soul}
\usepackage{booktabs}
\usepackage{changes}
\usepackage{braket}
\usepackage{physics}
\usepackage{tabularx,multirow}

\begin{abstract}
The significance of Mendeleev's periodic table extends beyond the classification of elements; it lies in its remarkable predictive power for discovering new elements and properties, revealing the underlying symmetrical patterns of nature that were only fully understood with the advent of quantum mechanics. Fundamental material properties, such as electron transport and magnetism, are also governed by crystal symmetry. In particular, spin transport depends on the spin polarization of electronic states, and recently discovered materials where the electron spin polarization is independent of momentum - a property known as a persistent spin texture (PST) - promise extended spin lifetime and efficient spin accumulation. In this paper, we establish the general conditions for the existence of symmetry-protected PST in bulk crystals. By systematically analyzing all 230 crystallographic space groups, similar to elements in the periodic table, we demonstrate that PST is universally present in all nonmagnetic solids lacking inversion symmetry. Using group theory, we identify the regions within the Brillouin zone that host PST and determine the corresponding directions of spin polarization. Our findings, supported by first-principles calculations of representative materials, open the route for discovering robust spintronic materials based on PST.

\end{abstract}

\begin{document}
\title{Universal symmetry-protected persistent spin textures in nonmagnetic solids}
\author{Berkay Kilic}
\affiliation{Zernike Institute for Advanced Materials, University of Groningen, Nijenborgh 3, 9747 AG Groningen, The Netherlands}
\author{Sergio Alvarruiz}
\affiliation{Zernike Institute for Advanced Materials, University of Groningen, Nijenborgh 3, 9747 AG Groningen, The Netherlands}
\author{Evgenii Barts}
\affiliation{Zernike Institute for Advanced Materials, University of Groningen, Nijenborgh 3, 9747 AG Groningen, The Netherlands}
\author{Bertjan van Dijk}
\affiliation{Zernike Institute for Advanced Materials, University of Groningen, Nijenborgh 3, 9747 AG Groningen, The Netherlands}
\author{Paolo Barone}
\affiliation{CNR-SPIN Institute for Superconducting and other Innovative Materials and Devices, Area della Ricerca di Tor Vergata, Via del Fosso del Cavaliere 100, I-00133 Rome, Italy}
\author{Jagoda S\l awi\'{n}ska}
\affiliation{Zernike Institute for Advanced Materials, University of Groningen, Nijenborgh 3, 9747 AG Groningen, The Netherlands}

\maketitle

\begin{figure*}
    \centering
    \includegraphics[width=\textwidth]
    {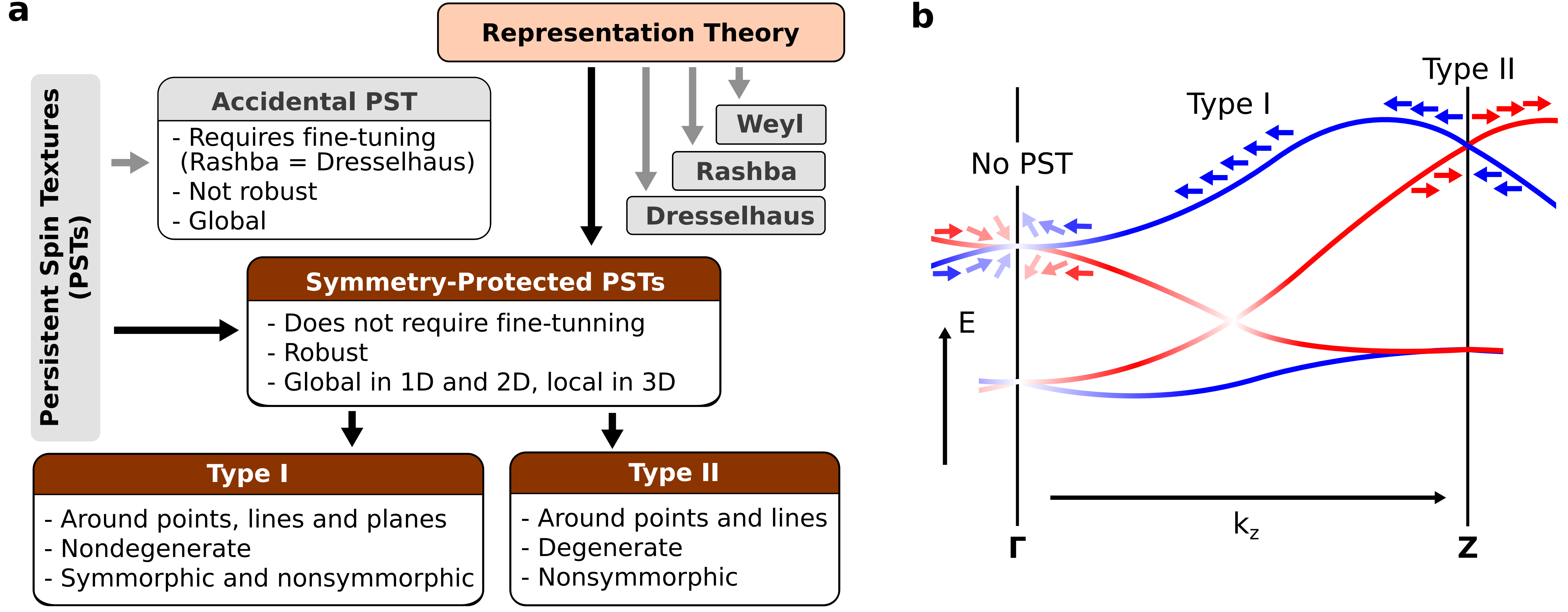}
    \caption{(a) Different types of spin textures in nonmagnetic bulk crystals that can be classified by the representation theory. We distinguish between accidental and symmetry-protected PSTs, as well as type I and type II SP-PST. (b) Schematic illustration of type I and type II in the presence of a single screw and time-reversal symmetries. Red and blue denote opposite spin polarization. Near the $\Gamma$ point, Kramers partners have eigenvalues with opposite signs, allowing nonzero perpendicular components of spin, whereas at the $Z$ point Kramers partners have eigenvalues with the same signs, preventing perpendicular spin components. Type I PST occurs around the single bands near the high-symmetry line. Type II PST occurs near the $Z$ point. 
    %The screw rotation also enforces a single symmetry-enforced band crossing, giving rise to an hourglass type band diagram with a nonsymmorphic Weyl node.
}
    \label{fig:table1}
\end{figure*}    

%In recent years, there has been growing interest in materials and structures with complex magnetic textures, such as spin spirals or skyrmions, which, beyond fundamental importance can be used in various applications \cite{ramesh}. Similar intricate textures can be found

In nonmagnetic materials with low crystal symmetry, strong spin-orbit coupling (SOC) gives rise to an effective magnetic field $B \sim \grad V(\bf{r}) \times \bf{p} $, where $\bf{p}$ is the momentum, and $V(\bf{r})$ is the crystal potential. The effective field interacts with the electron spin $\sigma$, causing spin-splitting and spin polarization of electronic states in $\bf{k}$-space \cite{zunger_splitting}. As a result, a variety of spin textures develop throughout the Brillouin zone (BZ) \cite{zunger_classification}, which beyond fundamental importance, play a crucial role for emerging spin-orbitronics applications \cite{analogs}. Spin textures are often categorized into three types based on their behavior around high-symmetry $\bf{k}$-points: Rashba (spin perpendicular to momentum) \cite{rashba_perspective}, Dresselhaus (a mix of perpendicular and parallel spin-momentum locking) \cite{dresselhaus}, and Weyl (parallel spin-momentum locking) \cite{schroeter}. These types stem from effective $\bf{k}\cdot\bf{p}$ models but extending them beyond the linear-in-$\bf{k}$ approximation for the systems with general symmetry remains a challenge \cite{diversity}. Moreover, particularly intriguing is the persistent spin texture (PST), where spin polarization remains unidirectional and independent of momentum \cite{schliemann}, falling outside the conventional classification. This unique property minimizes spin dissipation and supports a long spin lifetime which is highly desirable for spintronics devices \cite{su2, evgenii}, and calls for a more general framework that can comprehensively describe spin textures in materials with any crystal symmetries. 
%materials with arbitrary crystal symmetries. 

%Such a classification would not only deepen our understanding of spin-related phenomena but also unlock new opportunities for designing advanced spintronic devices.
%Moreover, a particularly intriguing concept of a persistent spin texture (PST), where the spin polarization stays unidirectional and independent of momentum \cite{schliemann}, falls outside conventional classifications. This unique spin behavior across the BZ minimizes spin dissipation and supports the formation of a persistent spin helix in real space — a symmetry-protected spin precession mode with an extended spin lifetime \cite{su2}. These features make PST. % Thus, PST is highly promising for long-range spin transport and sought after for spintronics devices \cite{perovskite_photonics}. 
%Unlike magnetic textures, spin textures in $k$-space enable electrical manipulation of spin degrees of freedom, making them suitable for low-power applications.

PSTs were originally discovered in two-dimensional (2D) semiconductor quantum wells, where the Rashba and Dresselhaus parameters were precisely balanced through adjustments in quantum well width, doping levels, and external electric fields \cite{koralek}. However, PSTs in these systems are sensitive to perturbations, and the need for precise parameter tuning, combined with temperature limitations, impedes their practical applications. Recently, Tao and Tsymbal showed that symmetries in crystalline solids can enforce uniform spin polarization of states in specific regions of the BZ, creating symmetry-protected persistent spin textures (SP-PSTs) \cite{tsymbal1}. These SP-PSTs are robust against symmetry-preserving perturbations because they do not depend on the microscopic details of the crystal Hamiltonian. However, only a small class of crystals with nonsymmorphic symmetries has been shown to support PSTs around certain high-symmetry points in the BZ \cite{tsymbal1, lingling, carmine, djani, snte, perovskite_photonics, BiNanoLett}. The general criteria for SP-PSTs have only been formulated for a subset of 2D materials \cite{2dpst}, and otherwise, potential candidates are identified based on the presence of specific symmetries that limit the range of available structures \cite{rondinelli_strain, rondinelli_matter}. As a result, PST materials remain exceedingly rare. 
%primarily due to the lack of general criteria for their existence.

In this Article, we use group theory analysis to establish the general conditions required for SP-PST. We discover that it is a universal property of all nonmagnetic crystals lacking inversion symmetry, occurring around specific high-symmetry $\bf{k}$-points, lines, or planes in the BZ. When bands near the Fermi level coincide with these regions, the material can support persistent spin helix states protected against perturbations by crystal symmetry. By identifying two types of SP-PSTs in degenerate and nondegenerate bands, we reveal the physical mechanisms behind their formation. In contrast to the previous studies that mostly relied on symmetry eigenvalue analysis for crystals with few symmetries, our approach is based on representation theory applied to 230 crystallographic space groups (SGs). Specifically, we use the irreducible corepresentations of double grey groups that take into account the full symmetry of crystals, encoding all the symmetry restrictions on the spin components of Bloch bands. Our approach is not only effective for crystals with many symmetries but can also be extended to predict all types of spin textures in nonmagnetic materials, which enables their systematic classification \cite{zunger_classification}. The applicability of our method is sketched in Fig. 1a, which also provides an overview of the different types of PST.

\textit{Group theory analysis.} The symmetry of the $\mathbf{k}$ point imposes constraints on electron spin polarization of Bloch states, $\psi_i$, following from the transformation rules of their spin matrix elements $\ev{\sigma_\alpha}_{ij}=\bra{\psi_i}\sigma_\alpha\ket{\psi_j}$ given by~\cite{oguchi1}:
\begin{equation}
\label{eqn:master}
    \ev{\sigma_\alpha'}_{ij} = \sum_{k,k'} {D}_{ik}(g){D}^{*}_{jk'}(g)\ev{\sigma_\alpha}_{kk'}.
\end{equation}
Here, the Pauli matrices $\sigma_\alpha$ (with $\alpha =x,y,z$) transform as an axial vector under the group element $g$: $\sigma_\alpha' = g\, \sigma_\alpha \, g^{-1}$. The matrix elements $D_{ij}(g) = \bra{\psi_i} g \ket{\psi_j}$ define group representation on the Bloch states, and $i, j, k, k'= 1,...,d$ where $d$ is the dimensionality of the representation at the $\mathbf{k}$ point.
This method is the direct counterpart of the selection rules in quantum mechanics, commonly used in spectroscopy to identify possible transitions. They are allowed when the tensor product of the representations, under which the initial/final state and the perturbation potential transform, contains scalar representations in its Clebsch-Gordan decomposition~\cite{tinkham}.

While one could directly extend the approach of Ref.~\cite{oguchi1} by using representations of point groups that describe invariant regions in the Brillouin zone, we instead use corepresentations proposed by Wigner to treat space group and time reversal operations on equal footing~\cite{tinkham}. 
This approach allows us to classify unidirectional spin textures in any nonmagnetic material, even in nontrivial cases where the $\mathbf{k}$ point is invariant under a point symmetry transformation combined with time reversal.
To this end, we employ the irreducible corepresentations (irreps) of double magnetic space groups of type II, called grey groups \cite{opechowski, bradley}, taken from the Bilbao Crystallographic Server~\cite{elcoro2017double, elcoro2021magnetic}. 
By applying Eq.\eqref{eqn:master} to all symmetry elements for a given irrep, we identify $\bf{k}$ points where only a single spin component is allowed. 

Our results for 230 space groups are presented in Supplementary Table S1. For each symmetry group, the table lists the regions in the BZ -- including high-symmetry points, lines, and planes -- where SP-PST is present, as well as the direction of the allowed spin polarization. 
The centrosymmetric space groups are included in the list for consistency, but the inversion symmetry combined with time reversal ensures double degeneracy of opposite spin states in the whole BZ, implying no spin texture. 
Surprisingly, all 138 noncentrosymmetric space groups, except for the trivial group $P_1$, have regions where PST occurs, indicating that this property is universal among nonmagnetic bulk crystals without inversion symmetry. 
%If the bands with SP-PST lie near the Fermi level of the material, they will significantly impact the spin generation and spin dynamics \cite{evgenii}. 

\begin{figure*}
	\centering
	\includegraphics[width=\linewidth]{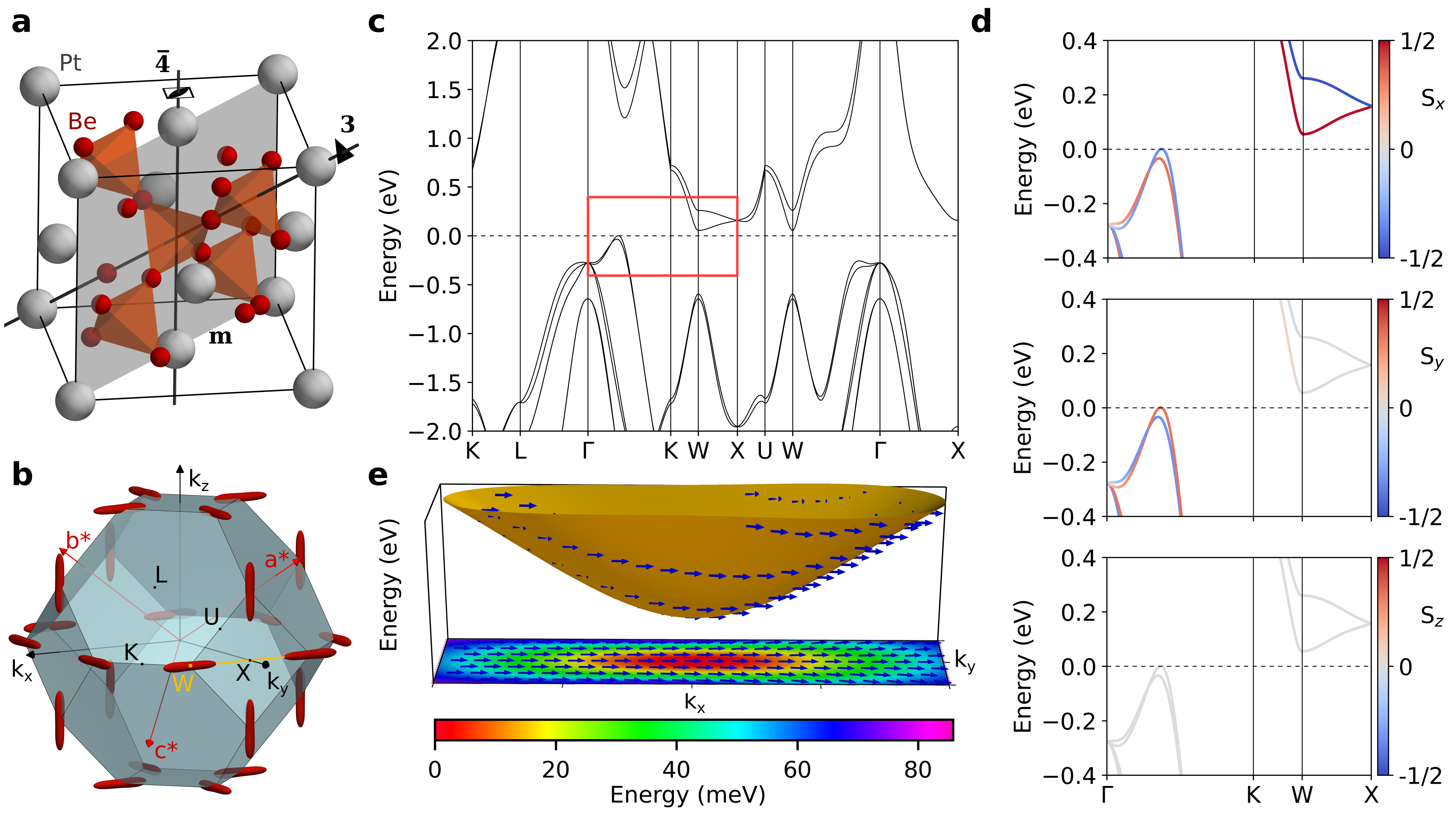}
	\caption{Electronic properties and persistent spin texture in Be$_5$Pt. (a) Crystal structure and representative symmetries. Be$_5$Pt has three kinds of symmetries: 4-fold inversion axes parallel to the conventional lattice vectors, 3-fold rotation axes parallel to the body diagonals, and mirror planes perpendicular to the face diagonals of the cube. (b) Brillouin zone with high-symmetry $k$-points and isoenergy pockets calculated at the energy $E = 50$ meV above the CBM. (c) The electronic structure calculated along the high-symmetry lines. The region marked by the red rectangle is magnified in panel (d) which additionally presents $S_x$, $S_y$, and $S_z$ components of the spin texture. (e) Zoom-in of the lowest conduction band around the $W-X$ line with the arrows representing the spin texture aligned along the $k_x$ axis. The color map shows the corresponding eigenvalues. 
    }
	\label{fig:be5pt}
\end{figure*}

We further categorize PSTs into two types: type I PST, which corresponds to a nondegenerate band, and type II PST -- to degenerate bands. 
For type I PST, Eq.~\eqref{eqn:master} reduces to $\ev{\sigma_\alpha'} = \ev{\sigma_\alpha}$. For example, if $g$ is a two-fold rotation around the $z$ axis, this equation becomes $(-\ev{\sigma_x},-\ev{\sigma_y},\ev{\sigma_z}) = (\ev{\sigma_x},\ev{\sigma_y},\ev{\sigma_z})$, enforcing PST with spin polarization parallel to the rotational axis, i.e.,  
$\ev{\sigma_{x,y}}=0$ and $\ev{\sigma_z} \neq 0$ (see Fig.\ref{fig:table1}(b))
Similarly, mirror symmetry enforces PST with spin polarization perpendicular to the mirror plane, and the presence of symmetries with different polarization axes constrains $\ev{\sigma_{x,y,z}}=0$. 
Nondegenerate bands only occur at $\bf{k}$ points that are not invariant under time-reversal symmetry, which excludes many points such as $\Gamma$, $X$, $Y$, and $Z$ in primitive lattices from supporting type I PST. We find that type I PSTs commonly occur in conventional and trigonal lattices where high-symmetry points can be not invariant under time-reversal symmetry.

Famous material realizations of type I PST are transition metal dichalcogenides, such as MoS$_2$ and WS$_2$, where the $K$ point in their honeycomb BZ displays robust out-of-plane PST, extending over a large region away from the $K$ point \cite{zeeman_effect, robust_zeeman}. Type I PST also occurs in bulk tellurium crystals along the vertical line connecting the $K$ and $H$ points, where the three-fold symmetry protects the PST \cite{npj_arunesh, karma_collinear}. Near these high symmetry points, spin canting arises from the admixture of other bands with different spin polarization. These bands can be included perturbatively, following the Luttinger-Kohn model, and therefore in low energy $\bf{k}\cdot\bf{p}$ expansion, stronger band separation in energy leads to persistent uniaxial spin polarization over a larger region in reciprocal space.

In type II PST, which involves degenerate bands, the analysis is more complicated. In general, even if uniaxial spin polarization is enforced in two degenerate bands, spin transfer in the vicinity of the band crossing can curl spins into a vortex-like spin texture, such as Rashba (or Dresselhaus). To avoid it in type II PST, the conditions to prevent spin flips or canting is that the irrep includes only one identity matrix $\pm \mathbb{I}_{d\times d}$ corresponding to a crystal symmetry different from the identity element of the group. 
%For the non-centrosymmetric space groups with time-reversal symmetry $d\leq 8$. 
%\hl{we ensure} that no inter-band matrix elements can cause spin flips or canting. The irrep includes \hl{only} an identity matrix $\pm \mathbb{I}_{d\times d}$, where $d\leq 8$ for the non-centrosymmetric space groups with time-reversal symmetry. 
This condition implies that the states forming the degeneracy must have the same eigenvalues for one of the symmetry elements in the little group. For instance, at the $\Gamma$ point in SG 4 ($P2_1$), the matrix of the two-fold screw rotation is 
\begin{equation}
\label{eqn:2sym}
    \Gamma(\Tilde{C}_{2z}) = \begin{pmatrix}
\pm i & 0 \\
0 &  \mp i 
\end{pmatrix},
\end{equation}
and by applying Eq.\eqref{eqn:master}, we find $\ev{\sigma_{x,y}}_{11,22} = 0$, $\ev{\sigma_z}_{11,22} \neq 0$, $\ev{\sigma_{x,y}}_{12,21} \neq 0$ and $\ev{\sigma_z}_{12,21} = 0$. These conditions lead to nonzero spin components in the $x$, $y$, and $z$ directions near the $\Gamma$ point. 
In contrast, at the $Z$ point, the matrix is
\begin{equation}
\label{eqn:2nsym}
    Z(\Tilde{C}_{2z}) = \begin{pmatrix}
\pm 1 & 0 \\
0 &  \pm 1 
\end{pmatrix},
\end{equation}
which results in $\ev{\sigma_{x,y}}_{ij} = 0$ and $\ev{\sigma_z}_{ij} \neq 0$ for all $i,j$, enforcing type II PST with spin polarization parallel or antiparallel to the screw rotation axis (see Fig. 1b). 

Symmetries with irreps of the form in Eq.~\eqref{eqn:2nsym} enforce PST along their invariant axes. Notably, such irreps can only occur in the presence of nonsymmorphic symmetries, where the fractional translation part of the symmetry operator allows for real eigenvalues at the BZ edges, permitting the diagonal matrices as in Eq.~\eqref{eqn:2nsym}. PST of this type was first proposed through an eigenvalue analysis of glide mirror symmetries in orthorhombic space groups~\cite{tsymbal1}. However, our approach provides a systematic generalization applicable to all possible symmetries and space groups. Besides, we discover that in orthorhombic crystals, there are more points and lines around which PST is symmetry-enforced, particularly along the edges of glide mirror planes (see Table S1). 

\begin{figure}
    \centering
    \includegraphics[width=\linewidth]{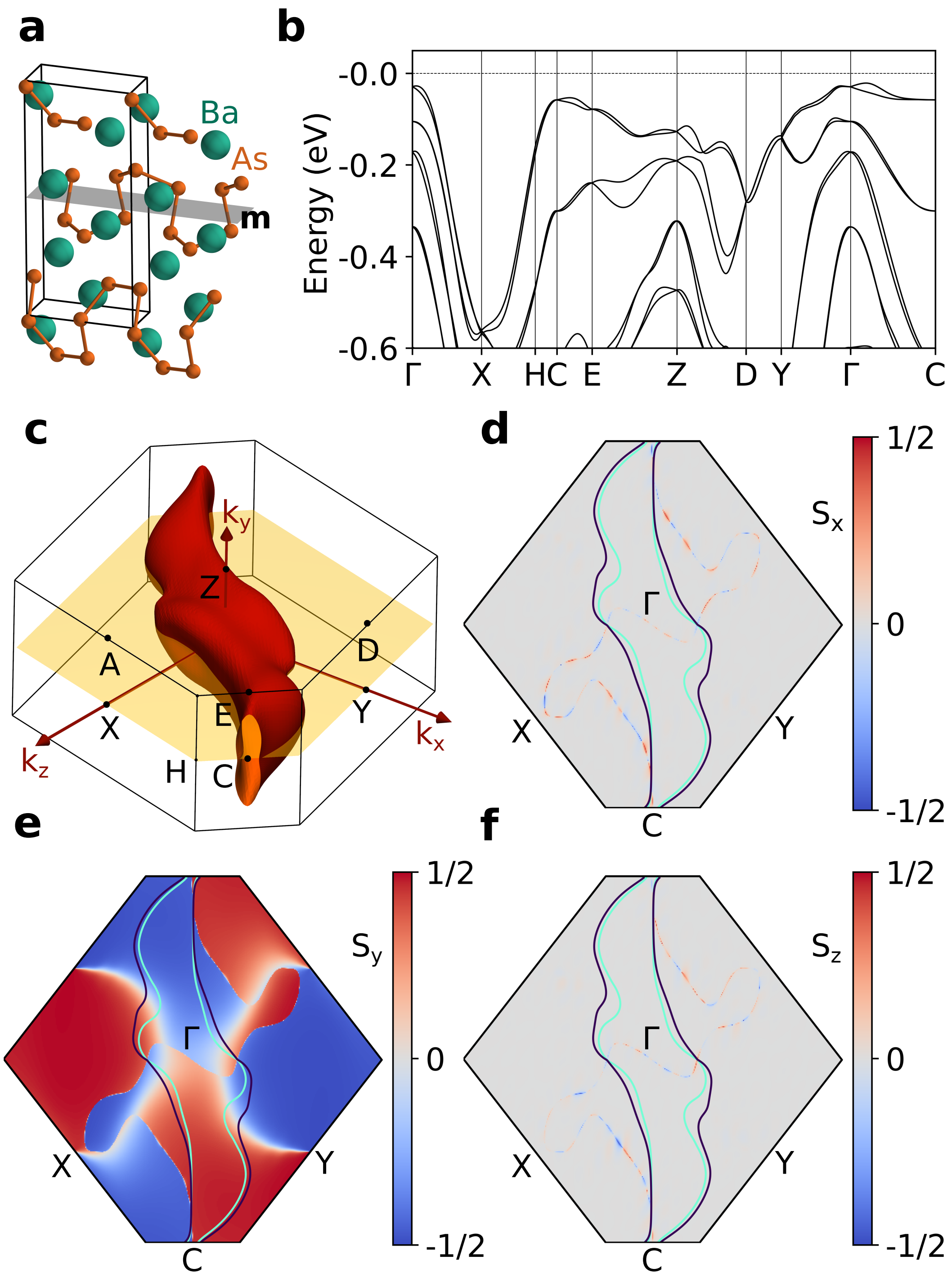}
    \caption{Geometry and electronic properties of BaAs$_2$. (a) Crystal structure with the mirror plane marked in grey. (b) Band structure along the high-symmetry path; the band gap is 0.34 eV. Only the valence bands are displayed. (c) Isoenergy surface at $E=-50$ meV. (d)-(f) Spin texture of the topmost valence band at the $\Gamma YX$ plane, marked as yellow in panel (c). The isoenergy contours correspond to $E=-50$ meV. }

    \label{fig:baas2}
\end{figure}   

\begin{figure*}
    \centering
    \includegraphics[width=\linewidth]{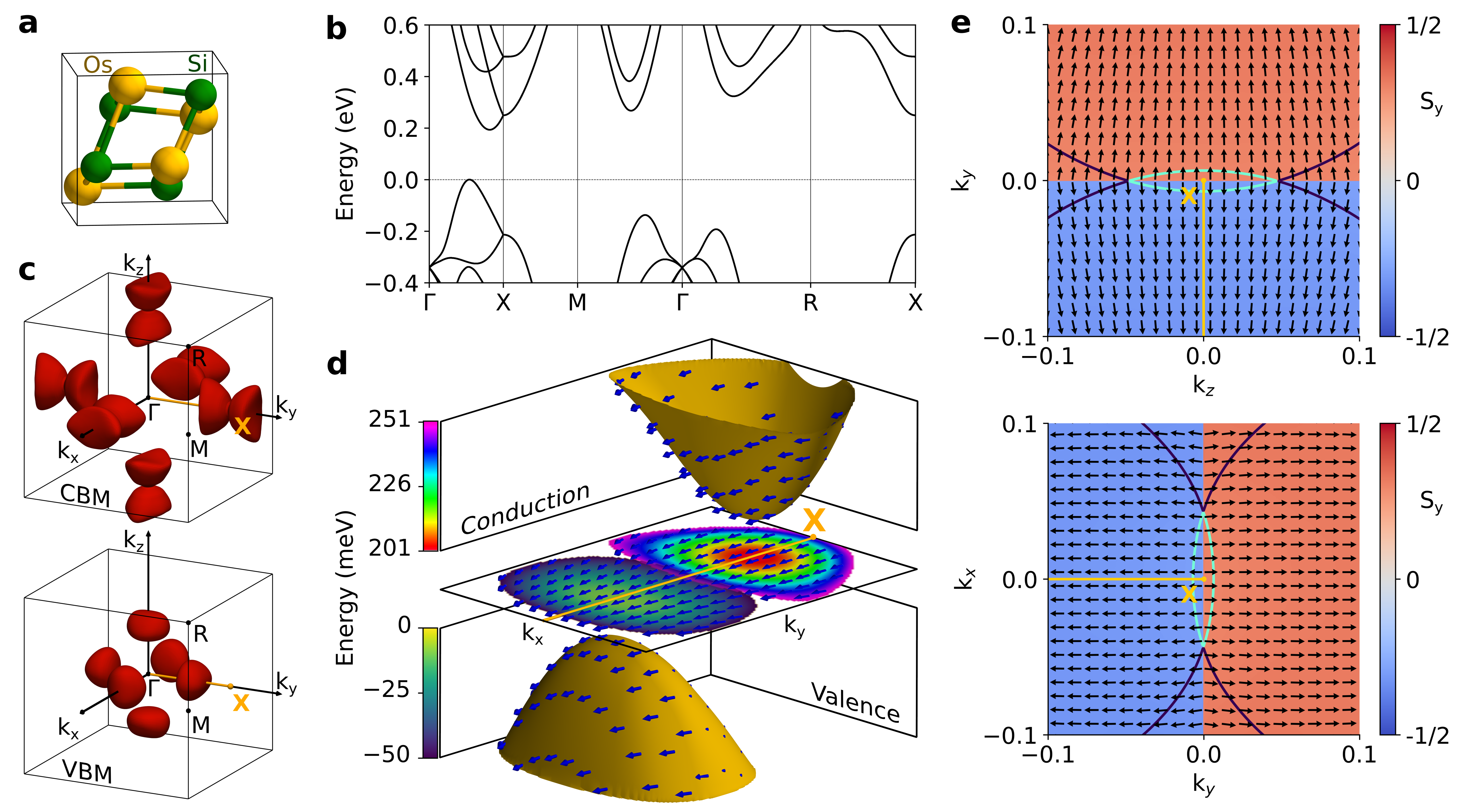}
    \caption{Chiral OsSi as an example of SP-PST of type II. (a) Cubic unit cell of OsSi. (b) Electronic structure along the high-symmetry lines; the energy gap is 200 meV. (c) Isoenergy surfaces of the lowest (highest) conduction (valence) bands. The energy levels were set to +50 meV and -50 meV with respect to CBM and VBM, respectively. (d) Three-dimensional view of the valence and conduction bands closest to the Fermi level. The arrows represent the spin texture. The central plane displays the maps of energy eigenvalues. The energy scale is in line with panel (b). (e) Maps of the spin texture at different planes. The wave vector is given with respect to the $X$ point which is set to zero. The isoenergy contours correspond to $E=50$ meV.}

    \label{fig:ossi}
\end{figure*}  

%Based on our symmetry analysis,
\textit{First-principles calculations.} To show the predictive power of our approach, we identified representative materials hosting different types of SP-PST. The initial search was conducted using the \textsc{aflow} materials database \cite{aflow}, which, despite containing nonrelativistic electronic structures, provided a useful first indication of whether PST-associated paths in the BZ have bands near the Fermi level. To guide our search, we selected materials with strong SOC to increase band splitting, and small unit cells to streamline the calculations. We then performed detailed density functional theory (DFT) calculations on a few promising candidates, with the computational details for each material described in the Methods section.

First, we focus on a cubic crystal Be$_5$Pt (SG 216) with type I SP-PST around a line. The material was previously synthesized, it has a large SOC and a small unit cell consisting of six atoms \cite{experimental_be5pt}. Its crystal structure and electronic properties are shown in Fig. \ref{fig:be5pt} and they agree well with the previous studies \cite{pseudogapped}. The conduction band minimum (CBM), located at the $W$ point, shows a huge spin-splitting of 205 meV, enabling applications well above room temperature. According to Table S1, Be$_5$Pt should have SP-PST of type I around the $W$ point and along the $XW$ line, with spins parallel to the $x$-direction in reciprocal space. This finding is fully confirmed by our spin-resolved band structures plotted in Fig. \ref{fig:be5pt}(d). Additionally, we explored the spin texture across the entire electron pockets near the CBM. For energy $E=+50$ meV, the pockets take on a cigar-like shape, growing larger at higher energies, though the spin texture remains well-aligned along the $x$ direction, with minor deviations in larger pockets (see Fig. \ref{fig:be5pt} (e)). Last, we estimated the spin Hall angle, finding it comparable to Pt (5-10$\%$, see Fig. S1 in SM). The combination of large spin-splitting, robust PST across the pockets, and a large spin Hall angle makes Be$_5$Pt promising for spintronics devices.
%\footnote{The calculated spin Hall conductivity close to the VBM is \sim 50 \unit, and charge conductivity of \sim 500 $ (\Omega\times $cm$)^{-1}$ \cite{pseudogapped}. This gives spin Hall angles of up to 10$\%$.}.
%

Figure \ref{fig:baas2} shows the calculated properties of BaAs$_2$ which crystallizes in a monoclinic lattice (SG 7). The structure consists of chiral chains of As separated by Ba cations. The only symmetry of the material is a mirror plane perpendicular to the $y$-axis. This symmetry implies type I SP-PST in the $\Gamma YX$ and $ZDA$ planes (see Table S1). Since BaAs$_2$ is a semiconductor with the VBM in the $\Gamma YX$ plane, it is convenient to analyze the spin texture of the valence bands. Near the Fermi level, the isoenergy surface consists of two sheets extending over the entire Brillouin zone along the $\Gamma YX$ plane; one of them is visualized in Fig. \ref{fig:baas2} (c). The spin texture of the topmost valence band is shown Fig. \ref{fig:baas2} (d)-(f), demonstrating SP-PST in the $y$ direction. Consistent with group theory predictions, the spin texture is unidirectional across the entire plane, except in a few regions that appear white in Fig. \ref{fig:baas2} (e), where the $S_x$ and $S_z$ components are also present. These regions correspond to the crossing with the second valence band. Because the $\Gamma YX$ plane hosts a type I PST, the band degeneracies disrupt PST. 

Finally, we calculated the properties of the chiral cubic crystal OsSi (SG 198), which belongs to the B20 materials class. Figure \ref{fig:ossi} shows its structural, electronic and spin properties. Both VBM and CBM lie in the $\Gamma-X$ line (see Fig. \ref{fig:ossi} (b)), and the bands near the Fermi level form several tiny pockets, as illustrated in Fig. \ref{fig:ossi} (c). The pockets close to VBM shown in Fig. \ref{fig:ossi} (d)-(e), feature type I PST along the $y$ axis. Moreover, the CBM is located near the X point, which exhibits type II PST also along the $y$ axis. The type II PST is enforced by the 2-fold screw rotation, which plays a similar role to the glide-mirror symmetries studied in \cite{tsymbal1}. 
%Despite the presence of additional symmetries with different polarization axes at X point, the real eigenvalues of the 2-fold screw symmetry form irreps of the \eqref{eqn:2nsym}, enforcing spin polarization along the $y$ axis.
Consequently, the band degeneracies do not disrupt the PST around this point, as shown in Fig. \ref{fig:ossi} (e). We believe that SP-PST will be crucial for sustaining long-range transport of spin density generated via the Rashba-Edelstein effect recently studied in this material \cite{karma_collinear}.

In summary, we demonstrated that persistent spin textures can be found in materials with any crystal symmetries, contrary to the common belief that they are an exotic property limited to a few special material classes. The exact knowledge of the PST location in the BZ for each space group allows for an easy design of materials with PST near the Fermi level, where they significantly influence the material properties and play a crucial role in spin generation and spin dynamics. Additionally, our theoretical approach based on representation theory not only allows for the classification of all types of spin textures in nonmagnetic solids but can also be extended to broader material classes, including altermagnets and other magnetic systems as well as centrosymmetric structures with hidden spin textures \cite{zunger_hidden, zunger_hidden_pst}. We believe that our findings represent a significant step toward identifying materials that combine strong spin-orbit coupling with extended spin lifetimes, addressing one of the most critical challenges in spintronics.

% \begin{table*}[ht]
%     \caption{}
%     \label{tab:ParamValues}
%     \begin{tabularx}{\textwidth}{|l|X|X|X|}
%         \hline
%         SG & Points & Lines & Planes  \\ 
%         \hline
%         
%          3, 4 & $\Gamma(s_z)$,$X(s_z)$, $Y(s_z)$,$Z(s_z)$, $C(s_z)$,$A(s_z)$, $D(s_z)$, $E(s_z)$ & $\Lambda(s_z)$,$U(s_z)$, $V(s_z)$,$W(s_z)$ & - \\
%  \hline
%  5 & $\Gamma(s_z)$,$Z(s_z)$, $Y(s_z)$,$L(s_z)$ & $\Lambda(s_z)$, $U(s_z)$ & - \\
%  \hline
%  6,7 &  $\Gamma(s_z)$,$X(s_z)$, $Y(s_z)$,$C(s_z)$, $Z(s_z)$,$A(s_z)$, $D(s_z)$,$E(s_z)$  & - & $F(s_z)$, $G(s_z)$ \\
%  \hline
%  8,9 & $\Gamma(s_z)$,$X(s_z)$, $Y(s_z)$ & - & $B(s_z)$ \\
%  \hline
%   &  & &  \\
%  \hline
%   & & &  \\
%  \hline
%     \end{tabularx}
% \end{table*}

\section*{Methods}
We performed first-principles calculations for all the materials using the Quantum Espresso package \cite{qe1, qe2}. The ion-electron interaction was treated using the fully relativistic projected-augmented wave pseudopotentials from the pslibrary database \cite{pslibrary} and the electron wave functions were expanded in a plane-wave basis using converged kinetic energy cutoff values. The exchange and correlation interaction was taken into account via the generalized gradient approximation (GGA) parameterized by the Perdew, Burke, and Ernzerhof (PBE) functional \cite{pbe}. The atomic coordinates of the structures were relaxed with the convergence criteria for energy and forces to $10^{-8}$ Ry and $10^{-4}$ Ry/bohr, respectively. The electronic convergence threshold was set to $10^{-9}$ Ry. The BZ sampling at the DFT level was performed following the Monkhorst-Pack scheme \cite{monkhorst-pack}, using $k$-meshes converged separately for each material. SOC was taken into account self-consistently in all the calculations except for the structural optimizations. The post-processing calculations of spin textures and spin Hall conductivity were performed using the \textsc{paoflow} code \cite{paoflow1, paoflow2}.

Be$_5$Pt (SG 216) was simulated in a cubic unit cell using the experimental lattice constant $a=5.97$ \AA\ \cite{experimental_be5pt}. We used the fixed occupations scheme and the $k$-points grids of $48\times 48\times 48$. The kinetic energy cutoff for wavefunctions was set to 280 Ry. 
BaAs$_2$ was calculated with a monoclinic cell with experimental lattice parameters $a=6.55$  \AA, $b=12.53$ \AA, $c=8.04$ \AA, $\beta$=127.75 \cite{experimental_baas2}. We used Gaussian smearing of 0.001 Ry and the $k$-points grids of $12\times 12\times 12$. The kinetic energy cutoff for wavefunctions was set to 110 Ry. OsSi was modeled in cubic unit cell with the experimental lattice constant $a=4.73$ \AA\ \cite{experimental_ossi}. We used the fixed occupations scheme and the $k$-points grids of $20\times 20\times 20$. The kinetic energy cutoff for wavefunctions was set to 110 Ry.

\section*{Acknowledgments}
\noindent 
%B.K. thanks Bertjan J.M. van Dijk for the support throughout the project. 
J.S. acknowledges the Rosalind Franklin Fellowship from the University of Groningen. We acknowledge the Dutch Research Council (NWO) for the grants NWA.1418.22.014 and OCENW.M.22.063. The calculations were carried out on the Dutch national e-infrastructure with the support of SURF Cooperative (EINF-8924) and on the H\'{a}br\'{o}k high-performance computing cluster of the University of Groningen.

\section*{Author contributions}
\noindent B.K. conceived and performed the representation theory-based symmetry analysis. S.A. performed first-principles calculations and prepared the figures. B.v.D. automatized the extraction of representation matrices and implemented the algorithm to solve the master equation. B.K., E.B. and J.S wrote the manuscript. All the authors contributed to the discussions and data analysis. J.S. initialized and supervised the project. 

\section*{Data availability}
\noindent The data associated with the manuscript will be available after the publication via DataverseNL. 

%\bibliography{ref.bib}

%apsrev4-2.bst 2019-01-14 (MD) hand-edited version of apsrev4-1.bst
%Control: key (0)
%Control: author (72) initials jnrlst
%Control: editor formatted (1) identically to author
%Control: production of article title (-1) disabled
%Control: page (0) single
%Control: year (1) truncated
%Control: production of eprint (0) enabled
%

\end{document}